# Cooperative and Interaction-aware Driver Model for Lane Change Maneuver

Jemin Woo, Changsun Ahn

*Abstract*— **To achieve complete autonomous vehicles, it is crucial for autonomous vehicles to communicate and interact with their surrounding vehicles. Especially, since the lane change scenarios do not have traffic signals and traffic rules, the interactions between vehicles need to be considered for the autonomous vehicles. To address this issue, we propose a cooperative and interaction-aware decision-making algorithm for autonomous vehicles that stochastically considers the future behavior of surrounding vehicles based on actual driving data. The algorithm is designed for both lane changing and lane keeping vehicles, and effectively considers interaction by using an interaction model based on relative information between vehicles with fewer states. To design the decision-making, the interaction model is defined as Markov decision process, and stochastic dynamic programming is used to solve the Markov decision process. We validate the effectiveness of our proposed algorithm in lane change scenarios that require interaction. Our results demonstrate that the proposed algorithm enables cooperative and interaction-aware decision-making while accommodating various driving styles. Additionally, by comparing it with other methods, such as the intelligent driver model and game theory-based decision-making, we validate the safety and comfortable decision-making of our proposed algorithm. Furthermore, through driving with a human-driven vehicle, it is confirmed that the proposed decision-making enables to cooperatively and effectively drive with humans.**

## I. INTRODUCTION

AUTONOMOUS vehicles are important role of intelligent transportation systems, and numerous studies have been conducted in this area [1-4]. While autonomous vehicle technologies have rapidly advanced in recent years, there remain several limitations that must be addressed to achieve fully autonomous vehicles. One such limitation is the need to consider the implicit interactions that occur between vehicles and reflect them into decision-making processes [5-7]. If these interactions are not properly accounted for, unexpected accidents may occur, which can lead to increased traffic congestion, higher collision risk, and reduced driving efficiency. For example, there have been reports of unexpected accidents involving autonomous vehicles, which have had negative impacts on energy consumption and overall safety [8, 9]. To ensure safe and efficient driving, autonomous vehicles must account for implicit interactions with other vehicles.

Implicit interactions between vehicles often occur in scenarios such as intersections [10, 11], roundabouts [12, 13], and lane changes [14-16]. Most intersections are equipped with traffic signals and governed by specific rules, enabling vehicles to often drive without requiring frequent interaction. Since the roundabouts do not have traffic signals, frequent interactions are required. Nevertheless, the roundabouts still have traffic rules such as priorities, which allows for driving without excessive interaction. In contrast, lane change scenarios rely primarily on implicit communication between vehicles, as there are no traffic signals or rules governing these situations. Consequently, interactions between vehicles are nearly inevitable during lane changes. This study focuses on developing a decision-making algorithm for autonomous vehicles in lane change scenarios.

There are many studies related to the decision-making for autonomous vehicles in lane change scenarios, and the decision-making algorithms have been developed using various methods such as support vector machines [17], predictive Stanley lateral controllers [18], and reinforcement learning [19]. However, these studies have largely focused on improving the efficiency and safety of the lane changing vehicle itself, without considering the interactions with surrounding vehicles. Similarly, current automated lane change technology in commercial vehicles is limited to very safe cases where there are no vehicles in the adjacent lane, and does not factor in interactions with other vehicles [20].

To consider the interaction between vehicle, several recent studies have focused on developing decision-making strategies for lane change scenarios that consider interactions between vehicles [14, 21, 22]. For example, one study proposed a game theory-based automatic lane change controller that considers interactions, but it has limitations in optimizing only longitudinal maneuvers and making binary decisions about lateral changes [21]. Another study introduced a game theoretic model predictive controller that estimates the aggressiveness of other vehicles to effectively consider interactions, but it also only optimizes longitudinal acceleration and lateral lane changes [14]. A third study proposed a lane merging strategy using game theory that optimizes both longitudinal and lateral motion. However, the decision-making in this study is limited by using fixed values for longitudinal acceleration and lateral velocity, resulting in overly simplistic maneuvers [22]. Most studies that consider

This research was supported by the National Research Foundation of Korea funded by the Ministry of Science and ICT (No. NRF-2022R1A2C1004894) (*Corresponding author: Changsun Ahn*).

The authors are with Pusan National University, Busan 46241, South Korea (e-mail: woojmn@gmail.com; sunahn@pusan.ac.kr)

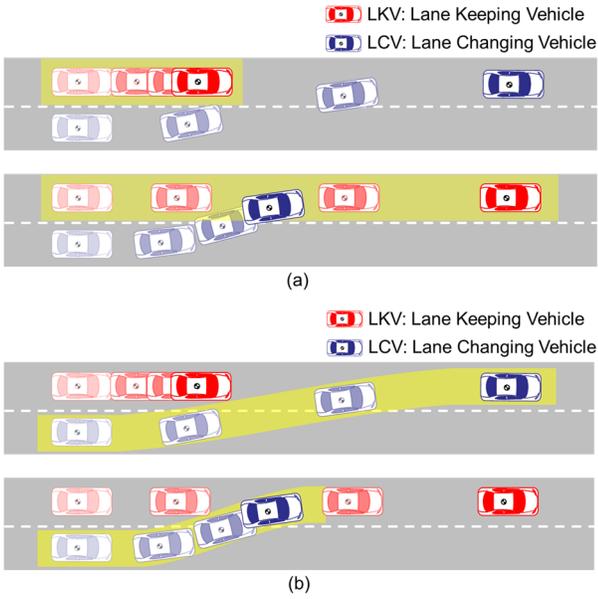

**Fig. 1.** Example of decision-making strategy for (a) LKV and (b) LCV.

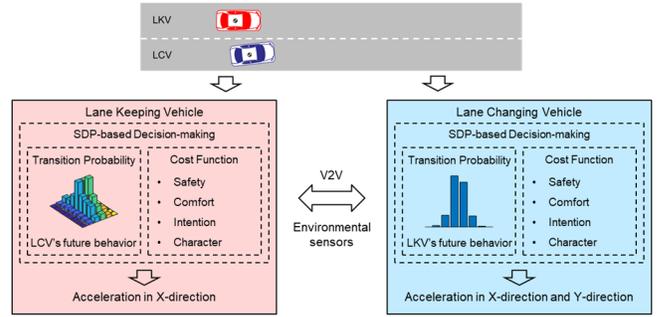

**Fig. 2.** Decision-making process for the LKV and LCV.

interactions between vehicles in decision-making use game theory as a framework. While game theory can effectively model interactions between multiple agents, it also has limitations. One of the limitations is it requires assumptions about the intentions and expected rewards of surrounding agents, which may be unknown information. As a result, it can be difficult to accurately reflect the intentions and rewards of actual other agents, and decision-making may be overly dependent on assumptions about opponent agents.

In this paper, we present a cooperative and interaction-aware decision-making algorithm for autonomous vehicles in lane change scenarios. The proposed algorithm stochastically takes into account the future behavior and intentions of other vehicles, and is developed for both a lane changing vehicle (LCV) and a lane keeping vehicle (LKV). To consider the future behavior of other vehicles, we use statistical driving data that is converted into probability models and expressed as a Markov process. We define the cooperative interaction model as a Markov decision process using relative information between vehicles. To effectively validate our proposed algorithm, we conduct simulations that involve the interactive driving of LKV and LCV. We evaluate the variation of speed according to each vehicle's characteristics and assess the safety and comfortable decision-making of our algorithm by comparing it with the intelligent driver model and a game theory-based decision-making algorithm. In addition, through driving with human-driven vehicles in an experimental environment, it is confirmed that the proposed decision-making can be driven cooperatively and interactively with humans.

The major contributions of this paper include in the following five items: 1) A cooperative interaction model is designed based on relative information between vehicles, allowing for effective interaction modeling with fewer states and enabling cooperative decision-making. 2) To the best of the authors' knowledge, no study has designed a decision-making algorithm considering interaction using stochastic dynamic programming (SDP). By incorporating actual driving data and probabilistically considering the future behavior of other vehicles, SDP enables effective interaction between vehicles and decision-making. 3) The proposed algorithm includes decision-making for LCV as well as LKV in lane change situations. 4) The proposed decision-making algorithm can interact with drivers using various opponent vehicles, including the intelligent driver model, game theory-based algorithms and human-driven vehicles. 5) The proposed algorithm is extensible beyond lane change scenarios to other driving situations.

The other parts of this paper are organized as follows: Chapter II formulates a target problem of this study, and Chapter III presents the proposed decision-making method. The cooperative and interaction-aware decision-making and the effectiveness of proposed decision-making is validated in Chapter IV. Chapter V provides an experimental validation of the proposed method while driving with a human-driven vehicle, and conclusions are drawn in Chapter VI.

## II. PROBLEM FORMULATION

This paper addresses the decision-making problem for autonomous vehicles in lane change scenarios, where implicit communications and interactions between vehicles often occur. Specifically, we focus on the interaction between a lane changing vehicle (LCV) and the nearest lane keeping vehicle (LKV) in an adjacent lane. During a lane change maneuver, the majority of lane keeping vehicles (LKVs) typically interact with only one LCV and, similarly, LCVs also interact with one LKV. Therefore, our study considers a scenario where there are two vehicles, LKV and LCV, on the road, as illustrated in Fig. 1. Other vehicles, apart from the LKV and LCV, are assumed to not interfere in the lane change situation.

Fig. 1 (a) shows an example of decision-making strategy for LKV based on the behavior of LCV, while Fig. 1 (b) illustrates an example of a decision-making strategy for LCV. During a lane change maneuver, the LKV needs to decide whether to yield and move behind the LCV or accelerate and move in front of the LCV. If the LCV desire to pass through an acceleration phase, the LKV should decelerate for the purpose of safety and efficient driving. Conversely, if the LCV yields during a deceleration phase, it is preferable for the LKV to accelerate for cooperative drive. Similarly, as shown in the Fig. 1 (b), the LCV should make a decision on whether to pass the LKV forward or yield and change lanes backwards, depending on the behavior of the LKV. These decisions are made based on implicit interactions considering the position, speed, and driving characteristics of the other vehicle. However, the drivers of each vehicle cannot accurately predict the future behavior, driving characteristics, and intentions of the opponent vehicle, leading to uncertainties in driving situations.

This paper deals with the interaction by stochastically considering the uncertain future behavior of the opponent vehicle. The schematic diagram of the decision-making design process for the LKV and the LCV is shown in Fig. 2. The decision-making algorithm is designed under the assumption that state information, such as the position and velocity of vehicles on the road, can be observed through environmental sensors and vehicle-to-vehicle (V2V) communication commonly used in autonomous vehicles [23-25]. The future behavior of the opponent vehicle is modeled as a probability distribution based on Markov chain, which is generated based on statistical driving data collected from actual roads. Therefore, the LKV considers the future behavior of the LCV as a probability model, and the LCV considers the future behavior of the LKV as a probability model. The cooperative interaction between the two vehicles is defined as a Markov decision process (MDP), and the cost function for the safety, comfort, intention, and character is designed to take into account the objectives of each vehicle. The decision-making algorithms for the LKV and LCV utilize the same framework, and the decision-makings of each vehicle expressed in MDP are solved using stochastic dynamic programming (SDP). For the simple optimal problem, we assume that the LKV does not include motion in the Y-direction. As a result, the output of the decision-making algorithm for LKV is the acceleration in the X-direction, while the accelerations in both X-direction and Y-direction are defined as the outputs of LCV.

## III. Cooperative and Interaction-aware Decision-making

This chapter presents a SDP-based cooperative and interaction-aware decision-making for autonomous vehicles. The proposed decision-making algorithm is specifically designed for lane keeping vehicle (LKV) and lane changing vehicle (LCV) in a lane change scenario, taking into account the individual objectives such as intentions, safety and characteristics etc. of each vehicle.

### A. Markov Chain-based Modelling Maneuver of Opponent Vehicle

As discussed in chapter II, the LKV and LCV drive while considering each other's presence in the lane change maneuver, and the future behavior of the opponent vehicle is uncertain information. Therefore, cooperative lane change maneuvering requires autonomous vehicles to incorporate the expected behavior of the opponent vehicle through interactive

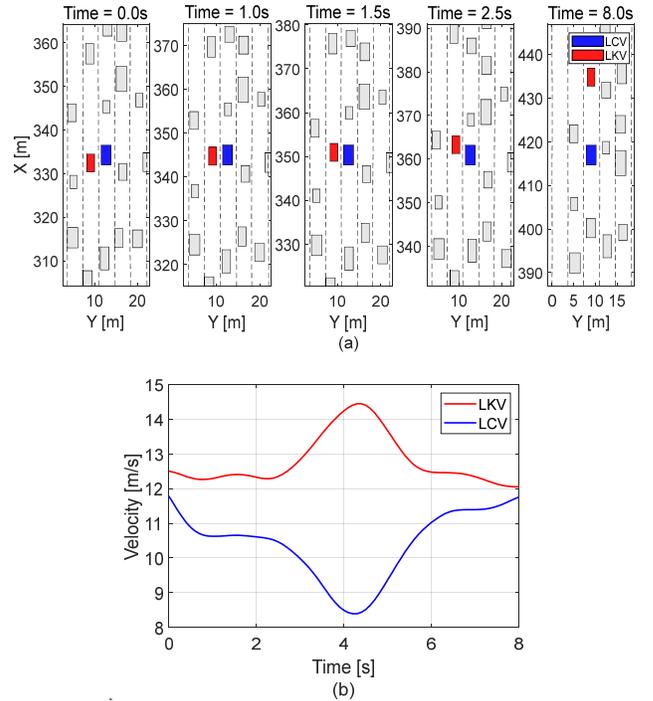

**Fig. 3.** Example of a scenario extracted from the NGSIM dataset (a) maneuver of the LKV and LCV and (b) velocity profiles.

decision-making. In this study, we propose a probabilistic approach that utilizes a Markov chain model to capture the uncertain future behavior of the opponent vehicle. The Markov chain model is defined based on relative global X-position $X^{rel}$, global Y-position $Y^{rel}$, and relative velocity in X-direction $v_X^{rel}$, which serve as the states of the LKV and LCV. Specifically, the Markov chain model for LKV represents the acceleration in X-direction $a_X^{LCV}$ and Y-direction of LCV $a_Y^{LCV}$, while the Markov chain model for LCV includes the acceleration in X-direction of LKV $a_X^{LKV}$, as follows:

$$
\begin{aligned}
p_{ijk,mn}^{LKV} = \Pr\{ & w^{LKV} = a_{X,m}^{LCV}, a_{Y,n}^{LCV} \mid \\
& X^{rel} = X_i^{rel}, Y^{rel} = Y_j^{rel}, v_X^{rel} = v_{X,k}^{rel} \}, \\
p_{ijk,q}^{LCV} = \Pr\{ & w^{LCV} = a_{X,q}^{LKV} \mid \\
& X^{rel} = X_i^{rel}, Y^{rel} = Y_j^{rel}, v_X^{rel} = v_{X,k}^{rel} \}, \\
\text{for } i = & 1, 2, \ldots, N_X, \quad j = 1, 2, \ldots, N_Y, \quad k = 1, 2, \ldots, N_v, \\
m = & 1, 2, \ldots, N_{a_X}, \quad n = 1, 2, \ldots, N_{a_Y}, q = 1, 2, \ldots, N_{a_X}
\end{aligned}
\tag{1}
$$

where $N_l$ ($l = X, Y, v, a_X,$ and $a_Y$) is the number of quantized grids, $\sum_{m=1}^{N_{a_X}} \sum_{n=1}^{N_{a_Y}} p_{ijk,mn}^{LKV} = 1$, $\sum_{q=1}^{N_{a_X}} p_{ijk,q}^{LCV} = 1$, and $p_{ijk,mn}^{LKV}$ and $p_{ijk,q}^{LCV}$ represent the one-step transition probability for the LKV and LCV. The relative X-position, Y-position, and relative velocity are computed as the differences between the corresponding states of LKV and LCV, and are defined as follows:

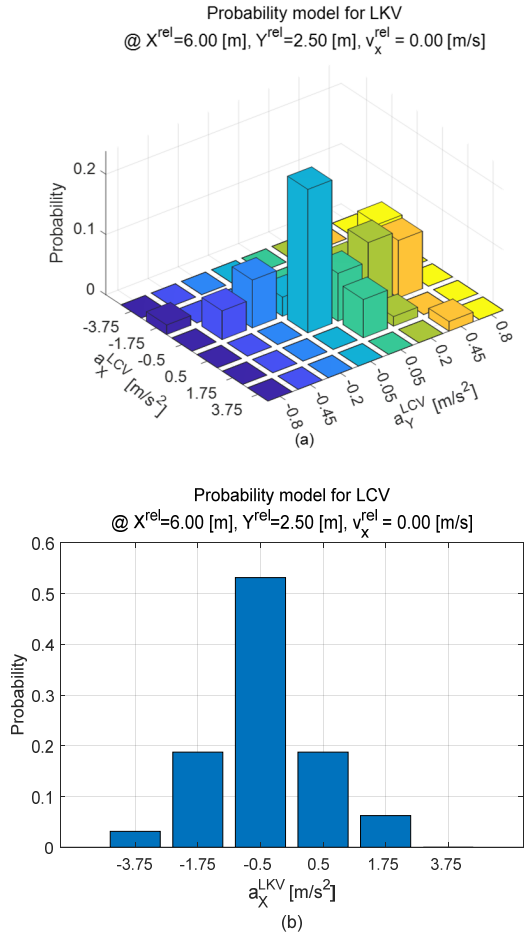

**Fig. 4.** Examples of transition probability models for (a) LKV and (b) LCV

$$X^{rel} = X^{LCV} - X^{LKV},$$
$$Y^{rel} = Y^{LCV} - Y^{LKV}, \qquad (2)$$
$$v_X^{rel} = v_X^{LCV} - v_X^{LKV},$$

where the superscript LCV denotes values corresponding to the state variables of LCV, while LKV represents values corresponding to the state variables of LKV.

The transition probability can be generally generated based on the driving historic data and standard driving cycle. This paper uses next generation simulation (NGSIM) open data to make the transition probability for the LKV and LCV [26]. The NGSIM dataset includes traffic data from a freeway road with many vehicles, encompassing various lane change scenarios. To effectively incorporate interaction-aware lane change scenarios into the development of the transition probability model, lane change scenarios are first extracted from the NGSIM data. Subsequently, only cases involving vehicle-to-vehicle interactions are utilized to generate the transition probability. As the determination of vehicle-to-vehicle interaction is subjective in nature, it is discerned based on the subjective judgment of the authors of this paper. Specifically, only cases where the speed change of the vehicle during lane change, as well as clear indications of intention such as yielding or overtaking, are considered during the extraction process. Fig. 3 illustrates an example of a scenario extracted from the NGSIM dataset. Fig. 3 (a) represents the maneuver of two vehicles, namely LKV and LCV, over time, while (b) depicts the velocity profiles of these two vehicles. The LCV decelerated in order to yield and change the lane, while the LKV interacted with the decelerating LCV by overtaking through acceleration.

Fig. 4 illustrates examples of transition probability models generated using the extracted scenario data. The probability model depicted in Fig. 4 (a) is utilized in the development of the decision-making algorithm for the LKV, while the model shown in Fig. 4 (b) is employed in the development of the decision-making algorithm for the LCV. The two probability models depict the probability values associated with a relative position $X^{rel}$ of 6m, a $Y^{rel}$ of 2.5m, and a relative velocity $v_X^{rel}$ of 0m/s. Consequently, the example scenario represents a situation where the LCV changes lanes while overtaking the LKV. In both of the probability models, the LKV has a high probability of deceleration, and the LCV has a high probability of acceleration. Therefore, the generated probability models accurately reflect natural driving patterns.

### B. Cooperative Interaction Model

The cooperative interaction model is needed to design for cooperative decision-making, and Markov decision process (MDP) is used to formulate the interaction model in this paper. The MDP is a useful framework for defining and solving optimal problems [27-29]. In order to formulate the MDP, the states and actions of the LKV and LCV are defined as follows:

$$s^i = [X^{rel}, \quad Y^{rel}, \quad v_X^{rel}, \quad v_Y^{rel}, \quad v_{intention}^i]^T,$$
$$\text{where } i = \{LKV, LCV\}, \qquad (3)$$
$$a^{LKV} = a_X^{LKV}, \quad a^{LCV} = [a_X^{LCV}, \quad a_Y^{LCV}]^T,$$

where $s$ represents the state of each vehicle, $a^i$ ($i = LKV, LCV$) is the actions of each vehicle, $v_Y^{rel}$ is relative velocity in Y-direction and $v_{intention}$ is intention of each vehicle, they are defined as follows:

$$v_Y^{rel} = v_Y^{LCV} - v_Y^{LKV},$$
$$v_{intention}^i = v_X^i - v_0, \text{ where } i = \{LKV, LCV\} \qquad (4)$$

The intention to follow the target speed, denoted as $v_{intention}$, is included as a state in the Markov decision process (MDP) to account for the intention of each vehicle to follow the target speed. This state is defined as the difference between the current vehicle speed and the target speed. This allows for flexibility in responding to various target speeds without the need for additional MDP solves. Furthermore, this paper assumes that the LKV does not move in the Y direction, so the LKV speed in the Y direction is assumed to be zero.

The next state ($s'$) is easily calculated by using the current state ($s$), actions, and transition probability model (TPM) and expressed as follows:

$$X_{k+1}^{rel} = X_k^{rel} + v_{X,k}^{rel} \cdot \Delta t$$
$$Y_{k+1}^{rel} = Y_k^{rel} + v_{Y,k}^{rel} \cdot \Delta t$$
$$v_{X,k+1}^{rel} = v_{X,k}^{rel} + \left( a_{X,k}^{LCV} - a_{X,k}^{LKV} \right) \cdot \Delta t \tag{5}$$
$$v_{Y,k+1}^{rel} = v_{Y,k}^{rel} + a_{Y,k}^{LCV} \cdot \Delta t$$
$$v_{intention,k+1}^{i} = v_{intention,k}^{i} + a_{X,k}^{i} \cdot \Delta t, \text{ where } i = \{LKV, LCV\}$$

where $\Delta t$ is a time step.

The state transition of each vehicle, which incorporates the actions of opponent vehicles, is utilized in the cooperative interaction model. The states of the interaction model are defined as relative information between vehicles. Unlike the general TPM that directly considers the probability of the next state given the current state and action, the proposed interaction model represents the future behavior of the opponent vehicle and calculates the next state based on this representation. The proposed interaction model effectively incorporates relative information and probabilistically accounts for the behavior of the opponent vehicle, allowing for a comprehensive consideration of vehicle-to-vehicle interaction and facilitating cooperative driving.

### C. Decision-making for LKV

In lane change driving, the driver of the LKV considers factors such as safety and comfort, and their driving intentions and characteristics are naturally reflected in their maneuvers. To develop the decision-making model for autonomous LKV, these factors are considered as cost functions of the MDP framework. The cost functions serve to represent the costs incurred during state transitions in the MDP, and are mathematically expressed as follows:

$$g^{LKV} = g_{safety}^{LKV} + g_{comfort}^{LKV} + g_{intention}^{LKV} + g_{character}^{LKV}, \tag{6}$$

where $g^{LKV}$ is a total cost of LKV, $g_{safety}^{LKV}$, $g_{comfort}^{LKV}$, $g_{intention}^{LKV}$, and $g_{character}^{LKV}$ represent the safety cost, comfort cost, intention cost, and character cost of LKV. Each cost function is defined based on the states and actions of the MDP, and is mathematically expressed as follows:

$$g_{safety}^{LKV} = 1 / \sqrt{\alpha_{safe,X} (X^{rel})^2 + \alpha_{safe,Y} (Y^{rel})^2},$$
$$g_{comfort}^{LKV} = \alpha_{comf} (a_X^{LKV})^2, \tag{7}$$
$$g_{intention}^{LKV} = \alpha_{int} (v_{intention})^2,$$
$$g_{character}^{LKV} = \alpha_{agg} X^{rel},$$

where $\alpha_i$ ($i = safe,X, safe,Y, comf, int$, and $agg$) is a tunable weighting factor. The safety cost is defined based on the relative distance, where larger relative distances are indicative of safer driving behavior. The comfort cost function is designed to minimize the square of acceleration, which serves as the input of the LKV because the driver typically experiences discomfort during vehicles acceleration and deceleration. The intention cost function is defined with the aim of minimizing the velocity difference between the target velocity. The character cost function is designed to incorporate the driver's individual driving characteristics, such as aggressiveness or conservativeness, into the decision-making model. Given that $X^{rel}$ is negative when the LKV is positioned ahead of the LCV, a higher value of $\alpha_{agg}$ would result in a more aggressive overtaking behavior, while a smaller $\alpha_{agg}$ would indicate a more conservative preference to stay behind the LCV. Therefore, the character of the LKV driver can be reflected through the tuning of $\alpha_{agg}$.

To solve the MDP and develop the decision-making model for the LKV, stochastic dynamic programming (SDP) is used in this paper. The SDP considers an expected total cost over an infinite horizon expressed as follows:

$$J_\pi^{LKV} (s^{LKV}) = \lim_{N \to \infty} \mathop{E}_{w_k^{LKV}} \left\{ \sum_{k=0}^{N-1} \gamma^k g^{LKV} (s_k^{LKV}, \pi(s_k^{LKV})) \right\}, \tag{8}$$

where $J_\pi^{LKV}$ indicates the expected total cost, $w^{LKV}$ represents the future acceleration of opponent vehicle LCV introduced in Markov process, $\gamma$ is the discount factor, and $\pi$ is the policy of the SDP. In this paper, the interaction model and cost functions are employed in the SDP framework. To solve the SDP, there are many methods such as policy iteration [30], value iteration [31], and linear programming [32]. Due to its faster convergence time compared to value iteration and linear programming, the policy iteration algorithm has been widely utilized in various research studies. Therefore, the policy iteration algorithm is employed in this paper to solve the SDP, and the resulting optimized policy is expressed as follows:

$$\pi^{*,LKV} (s^{LKV}) =$$
$$\arg \min_{a^{LKV}} (g^{LKV} (s^{LKV}, a^{LKV}) + \mathop{E}_{w^{LKV}} \left\{ \gamma J_\pi^{LKV} (s^{',LKV}) \right\}), \tag{9}$$

The optimized policy is utilized for the output of decision-making for LKV, as mathematically expressed as follows:

$$a_X^{*,LKV} = \pi^{*,LKV} (s^{LKV}). \tag{10}$$

### D. Decision-making for LCV

To develop the decision-making model for autonomous LCV, the factors mentioned in the LKV are equally considered as cost functions in the MDP framework, and are mathematically expressed as follows:

$$g^{LCV} = g_{safety}^{LCV} + g_{comfort}^{LCV} + g_{intention}^{LCV} + g_{character}^{LCV}, \tag{11}$$

where $g^{LCV}$ is a total cost of LCV, $g_{safety}^{LCV}$, $g_{comfort}^{LCV}$, $g_{intention}^{LCV}$, and $g_{character}^{LCV}$ represent the safety cost, comfort cost, intention cost, and character cost of LCV. The cost functions of LCV are defined similarly to the LKV, with the additional consideration of intention to change the lane, and expressed as follows:

$$
\begin{aligned}
g_{safety}^{LCV} &= 1 / \sqrt{\beta_{safe,X} (X^{rel})^2 + \beta_{safe,Y} (Y^{rel})^2}, \\
g_{comfort}^{LCV} &= \beta_{comf,X} (a_X^{LCV})^2 + \beta_{comf,Y} (a_Y^{LCV})^2, \\
g_{intention}^{LCV} &= \beta_{int} (v_{intention})^2 + \beta_{LC} (Y^{rel})^2, \\
g_{character}^{LCV} &= \beta_{agg} (-X^{rel}),
\end{aligned}
\tag{12}
$$

where $\beta_i$ ($i = safe,X$, $safe,Y$, $comf,X$, $comf,Y$, $int$, $LC$ and $agg$) is tunable weighting factor. The cost function for comfort is formulated to minimize the square of accelerations, which are two actions of the LCV. Additionally, the intention cost is adjusted by incorporating an additional cost term to ensure that the relative Y-position becomes zero for changing a lane. The cost function for character is adjusted to have the opposite sign of the LCV cost function, taking into consideration that the driver's behavior is considered aggressive when the weighting factor $\beta_{agg}$ is positive and conservative when the weighting factor is negative. Considering that $X^{rel}$ is positive when the LCV is positioned in front of the LKV, a higher value of $\beta_{agg}$ would suggest a more aggressive overtaking behavior, whereas a smaller $\beta_{agg}$ would imply a more conservative preference to stay behind the LKV. Therefore, the tuning of $\beta_{agg}$ can be used to reflect the various character of the LCV driver.

To solve the MDP and develop the decision-making model for the LCV, the SDP is used and the expected total cost over infinite horizon expressed as follows:

$$
J_\pi^{LCV}(s^{LCV}) = \lim_{N \to \infty} \underset{w_k^{LCV}}{E} \left\{ \sum_{k=0}^{N-1} \gamma^k g^{LCV}(s_k^{LCV}, \pi(s_k^{LCV})) \right\}, \tag{13}
$$

where $J_\pi^{LCV}$ is the expected total cost and $w^{LCV}$ represents the future acceleration of opponent vehicle LKV introduced in Markov process. To solve the SDP, the policy iteration algorithm is used similar to the LKV and the optimized policy is expressed as follows:

$$
\begin{aligned}
\pi^{*,LCV}(s^{LCV}) = \\
\arg\min_{a^{LCV}}(g^{LCV}(s^{LCV}, a^{LCV}) + \underset{w^{LKV}}{E} \left\{ \gamma J_\pi^{LCV}(s^{.LCV}) \right\}).
\end{aligned}
\tag{14}
$$

The optimized policy is used for the output of decision-making for the LCV, as mathematically expressed as follows:

$$
a_X^{*,LCV}, a_Y^{*,LCV} = \pi^{*,LCV}(s^{LCV}). \tag{15}
$$

The optimal policy of the LKV and LCV is the function of each states of MDP, can be utilized as a look-up table for the decision-making model. Therefore, the SDP-based decision-making algorithm has low computational complexity, making it suitable for implementation.

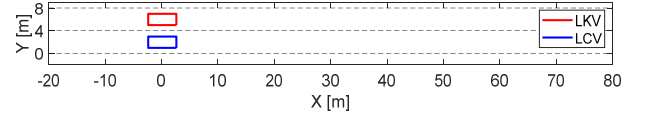

**Fig. 5.** Initial conditions for validation in a lane change scenario.

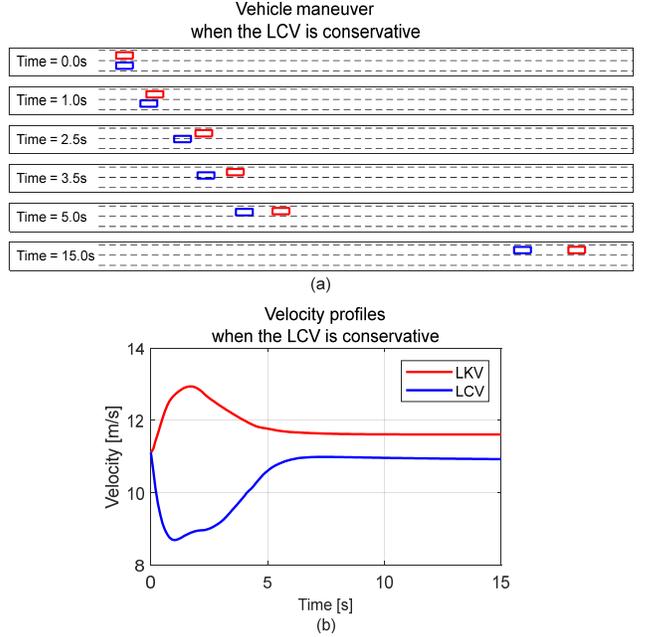

**Fig. 6.** Results of decision-making for the LKV with a conservative LCV (a) maneuver of two vehicles and (b) velocity profiles.

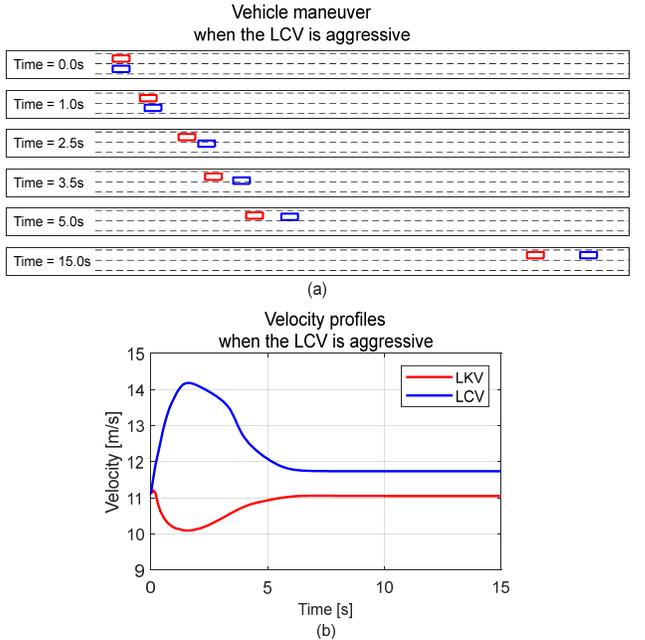

**Fig. 7.** Results of decision-making for the LKV with an aggressive LCV (a) maneuver of two vehicles and (b) velocity profiles.

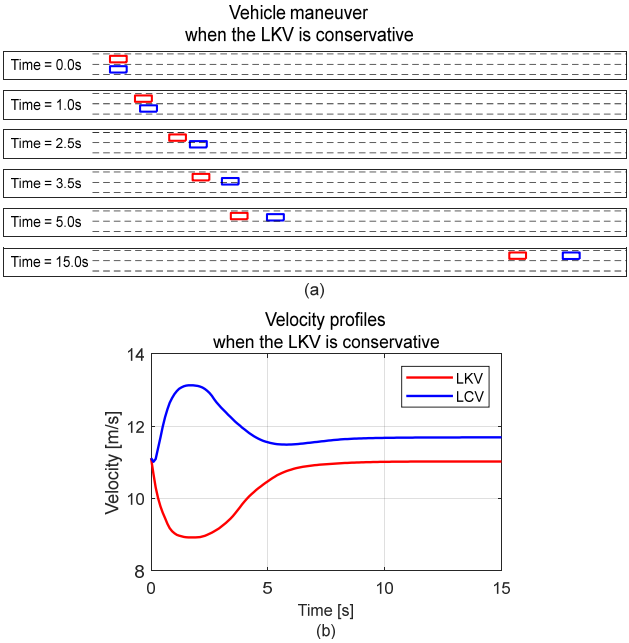

**Fig. 8.** Results of decision-making for the LCV with a conservative LKV (a) maneuver of two vehicles and (b) velocity profiles.

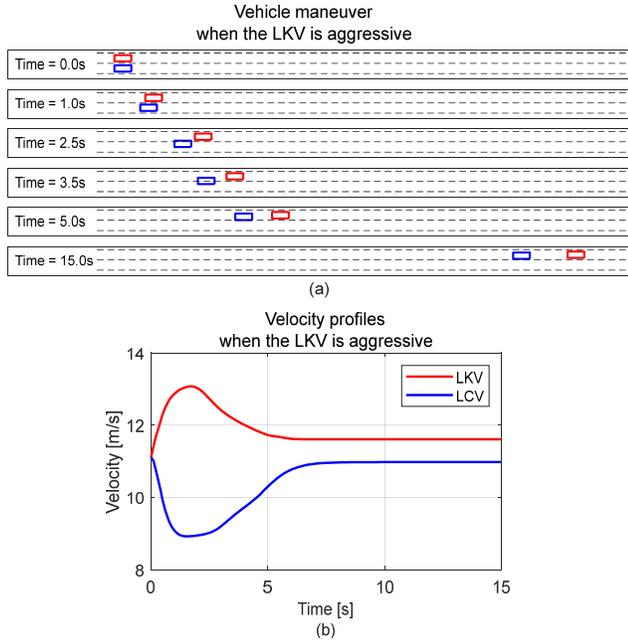

**Fig. 9.** Results of decision-making for the LCV with an aggressive LKV (a) maneuver of two vehicles and (b) velocity profiles.

## IV. VALIDATION

In this chapter, we evaluate the performance of the proposed decision-making algorithm for the LKV and LCV. This paper aims to demonstrate the cooperative driving and interaction-awareness of the proposed algorithm for the LKV and LCV by incorporating the interaction between vehicles and considering the driver's character. To validate the safe and comfortable decision-making, the proposed algorithm is compared with other decision-making algorithm. Furthermore, an experimental validation conducted in the driving with human-driven vehicles confirms that the proposed algorithm can be driven cooperatively through interaction with humans exhibiting various characteristics.

In order to validate the cooperative and interactive decision-making algorithm, the lane change scenario is initialized with conditions as depicted in Fig. 5, which facilitate implicit interactions. Additionally, the target speeds of both vehicles are set to match their initial speeds, and we only consider a situation where the LCV should change the lane. In the context of lane change driving where the initial X-position and speed are crucial, since it is impossible to change the lane at the same X-position and velocity of the LKV and LCV, the two vehicles should choose between overtaking or yielding. Thus, these initial conditions provide an effective means to verify the performance of the cooperative and interaction-aware decision-making algorithm.

### A. Evaluation of Cooperative and Interaction-aware Decision-Making

To validate the LKV decision-making algorithm, the policy for the LKV is kept fixed and two driving situations are considered: one where the LCV is a conservative, and another where the LCV is an aggressive. The LCV is driven based on the SDP-based decision-making algorithm and the character of the LCV is generated by tuning the weighting factor of the character cost.

The validation results of the LKV with a conservative LCV are depicted in Fig. 6, (a) presents the behavior of two vehicles according to time, and (b) displays the velocity over time. The findings confirm that the LKV demonstrates interactive response through acceleration as the conservative LCV undergoes deceleration and lane change maneuvers. Furthermore, it is observed that both vehicles converge to the target speed after completing the lane change interaction.

Fig. 7 shows the results of the LKV with an aggressive LCV. The findings confirm that the LKV demonstrates interactive response through deceleration, while the aggressive LCV accelerates and changes the lane. Furthermore, after completing the lane change and interaction, both vehicles drive to follow the target speed.

To validate the LCV decision-making algorithm, similar to LKV, the policy for the LCV is kept fixed and two driving situations are considered: one where the LKV is a conservative, and another where the LKV is an aggressive. The LKV is driven based on the SDP and the character is generated by tuning the weighting factor of the character cost.

Fig. 8 shows the results of the LCV with a conservative LKV. The LCV exhibits interactive driving behavior, accelerating while the conservative LKV undergoes deceleration. Fig. 9 shows the results of the LCV with an aggressive LKV. The findings confirm that the LCV demonstrates interactive response through deceleration while the aggressive LKV accelerate. Furthermore, in the both situations, after completing the lane change and interaction, both vehicles drive to follow the target speed.

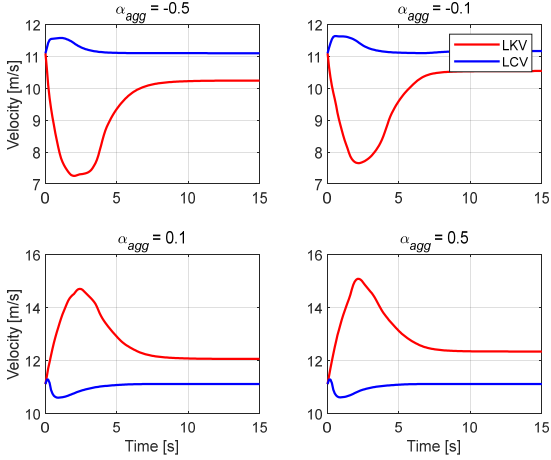

**Fig. 10.** Velocity profiles according to the variation of LKV's characteristics.

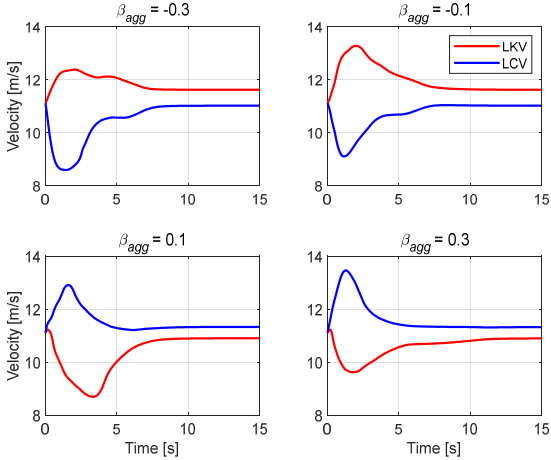

**Fig. 11.** Velocity profiles according to the variation of LCV's characteristics.

In both cases of the LKV and LCV, each vehicle drives through the interaction with the opponent vehicle. When the LKV experiences deceleration or acceleration, the LCV is capable of driving through acceleration and deceleration. If the LCV change the lane through the acceleration or deceleration, the LKV interactively drives by decelerating or accelerating. These results indicate that both vehicles can effectively interact without unnecessary large accelerations. As a result, the proposed decision-making algorithm has the potential to improve driving efficiency and enable cooperative driving on the road.

### B. Evaluation of Scalability of Driver Characteristics

As mentioned in chapter III, the proposed decision-making algorithm has the capability to incorporate various driver characteristics by adjusting the tunable factors. For example, when the weighting factor for safety is increased, the decision-making prioritizes increasing the distance from the opponent vehicle, resulting in safer decisions. If the weighting factor for intention is increased, the decision-making focuses on quickly returning to the initial speed for the LKV and LCV and leading to faster lane changes for the LCV. In particular, in the lane change scenarios, the lane change driving involves different types of drivers, including aggressive and conservative drivers. The aggressive drivers desire to overtake opponent vehicles through acceleration, while conservative drivers aim to yield through deceleration. This section demonstrates that the proposed decision-making algorithm can effectively capture various driver characteristics by adjusting the representative example's driver characteristics.

**TABLE I**
**INITIAL CONDITIONS OF SAFETY AND COMFORT TEST**

|  | LKV | LCV |
|---|---|---|
| X-position [m] | U(-5, 5) | U(-5, 5) |
| Velocity in X-direction [kph] | U(32.8, 47.2) | U(32.8, 47.2) |

The results depicting the variation of LKV's characteristics are presented in Fig. 10. The decision-making for the LKV is generated through various the tunable weighting factors of character cost. To verify the effectiveness of tuning these factors for the LKV, simulations are conducted with the decision-making for the LCV fixed by using same policy. A negative value of $\alpha_{agg}$ represents a conservative decision, while a positive value signifies an aggressive decision. It has been confirmed that smaller values of $\alpha_{agg}$ lead to more deceleration decision, while larger values of $\alpha_{agg}$ make to increase acceleration in the LKV.

The results illustrating the variation of LCV's characteristics are shown in Fig. 11. Similar to the validation for LKV, the simulation is conducted with a fixed decision policy for the LKV, and the effectiveness with respect to changing the weighting factor of character cost for the LCV are analyzed. It has been confirmed that smaller values of $\beta_{agg}$ result in more conservative decision-making through increased deceleration, while larger values of $\beta_{agg}$ lead to more aggressive driving behavior in the LCV through increased acceleration in the LKV. Therefore, the proposed algorithm has the capability to generate diverse decision-making models that reflect the characteristics of various drivers.

### C. Evaluation of Safety and Comfort

To evaluate safety and comfort of the decision-making, we compare the proposed algorithm with other decision-making algorithms. Specifically, we compare the proposed algorithm with the widely-used intelligent driver model (IDM) for interactive lane keeping vehicle, and game theory (GT)-based algorithm used for the lane changing vehicle decision-making in lane change scenarios. We analyze the collision rate between LKV and LCV to evaluate the safety, and the longitudinal jerk to evaluate the comfortable decision-making. To ensure the performance of each decision-making algorithm based on the initial position and velocity of the vehicles, the initial position and velocity of each vehicle are randomly set, and the simulation is repeated 1000 times. The initial conditions are listed in Table I, and U represents the uniform distribution. Furthermore, to assess the robustness for safety of the proposed algorithm, the collision rate performance is evaluated under both nominal conditions and harsh conditions. The nominal conditions represent situations where each vehicle has completely access to all the required information for decision-making, while the harsh conditions consider near-collision with noisy sensor.

TABLE II
Test Pairs of Decision-making Algorithms for Safety and Comfort Validation

|  | LKV | LCV |
|---|---|---|
| Test 1 | IDM | GT |
| Test 2 | IDM | SDP |
| Test 3 | SDP | GT |
| Test 4 | SDP | SDP |

TABLE III
Results of Validation for Proposed Decision-making in Test 1 vs. Test 4

|  | LKV / LCV | LKV / LCV |
|---|---|---|
| Algorithms | **SDP / SDP** | IDM / GT |
| Average of velocity [m/s] | **10.86 / 11.56** | 10.70 / 10.92 |
| Minimum of longitudinal jerk [m/s³] | **-26.97 / -19.98** | -50.75 / -100 |
| Maximum of longitudinal jerk [m/s³] | **23.63 / 22.21** | 51.34 / 100 |
| Collision rate [%] | **0** | 61 |
| Average of distance [m] | **21.09** | 11.64 |

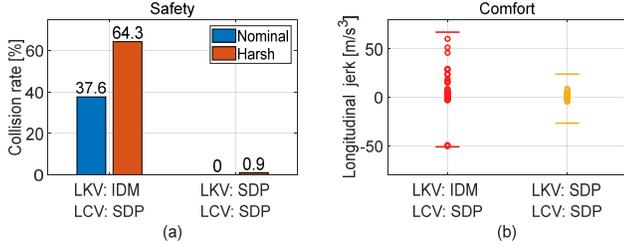

**Fig. 12.** Results of LKV validation in Test 2 vs. Test 4 (a) safety and (b) comfort under nominal conditions.

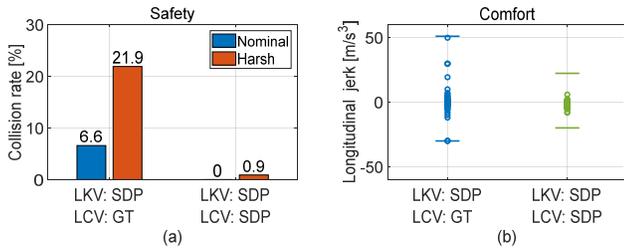

**Fig. 13.** Results of LCV validation in Test 3 vs. Test 4 (a) safety and (b) comfort under nominal conditions.

The tests consist of four possible combinations, as illustrated in Table II. To verify the safe and comfortable decision-making for the LKV, we compare the proposed LKV algorithm with the IDM, assuming that the LCV is driven using the same SDP-based decision-making algorithm. This comparison is conducted between Test 2 and Test 4. The IDM is designed to have the same maximum acceleration as the proposed algorithm, and its control variable is the longitudinal acceleration, expressed as follows:

$$\dot{v}_x = a_{\max} \left\{ 1 - \left( \frac{v_x}{v_{x,0}} \right)^{\delta} - \left( \frac{s^*(v_x, \Delta v_x)}{s_\alpha} \right)^2 \right\}, \quad (16)$$

where $a_{max}$, $v_{x,0}$, $\delta$, $\Delta v_x$, $s^*$, $s_\alpha$ are the maximum acceleration, target speed, constant acceleration component, velocity difference, desired distance, and the minimum distance. To simulate the lane change scenario and validate the safe and comfortable decision-making for LKV, an interactive LCV is required. Therefore, the LCV is designed to drive using a fixed policy based on SDP.

The results of comparing SDP-based LKV and IDM-based LKV are presented in Fig. 12. The IDM-based LKV, which considers only longitudinal vehicles, exhibits a high collision rate of about 37.6%. In contrast, the SDP-based LKV demonstrates high safety with no collisions, even when the initial position and speed are changed. Furthermore, under the harsh conditions, the collision rate of the IDM-based LKV is increased by about 27.6%, whereas the SDP-based LKV shows a low increase of about 0.9%. Therefore, the proposed decision-making has robust performance for safety with respect to the harsh conditions.

In the longitudinal jerk results, the line represents maximum and minimum value of each algorithm, while the circles denote the sampled data from the 1000 repeated simulation. The results reveal that the IDM-based LKV algorithm has higher maximum and minimum jerk values than the SDP-based algorithm, and also exhibits higher variance of the jerk. Consequently, the proposed SDP-based algorithm can make more comfortable decision.

To validate the safe and comfortable decision-making for LCV, the proposed LCV algorithm is compared with the GT-based decision-making for LCV, assuming that the LKV is driven using the same SDP-based decision-making algorithm. This comparison is conducted between Test 3 and Test 4. The GT is a powerful framework to solve interaction between decision makers. One of the interesting papers showed game theory-based controller design for automatic lane changing [21]. The paper used the Stackelberg game for the longitudinal motion and PID controller for the lateral motion. We design a decision-making algorithm based on the paper [21], a payoff function of Starckelberg game is designed for safety and space requirement, and expressed as follows:

$$U_{total}(a_x) = f_w(a_x) \cdot$$
$$\left( (1 - \beta(q)) \cdot U_{safety}(a_x) + \beta(q) \cdot U_{space}(a_x) + 1 \right) - 1, \quad (17)$$

where $U_i$ ($i$=total, safety, space) represents the payoff function, $a_x$ denotes the future acceleration of the vehicle, $f_w(\cdot)$ is the penalty on the change of acceleration, $q$ is the aggressiveness of LCV, and $\beta(\cdot)$ is the cumulative distribution function, $\beta(\cdot)$ works as a weighting factor and is defined as a function of aggressiveness, $q$. The large value of aggressiveness means a high intention to overtaking and makes $\beta$ close to 1. The safety factor, $U_{safety}$, is designed to decrease to a negative value as the time headway approaches zero, indicating that safety is high when the time headway is sufficient. The space payoff, $U_{space}$, is defined as a function of the relative position and is designed to increase as the longitudinal distance margin increases, indicating that overtaking is more beneficial than yielding when there is enough space for changing the lane.

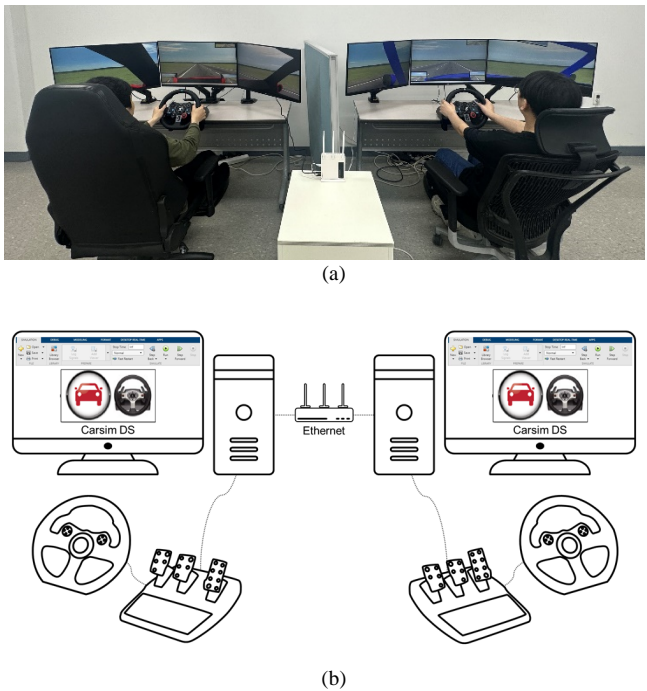

**Fig. 14.** Experimental environment (a) Networked traffic simulator for experimental validation and (b) System configuration.

The GT-based decision-making calculates the optimal longitudinal acceleration that maximizes the total payoff in (17). In addition, for the driver model of lateral motion, we also used a PI-based controller with a yaw rate reference, which has been shown to produce more naturalistic maneuvers than a simple position feedback controller, as demonstrated in the paper.

Fig. 13 presents the results of a comparative analysis between the proposed SDP-based LCV and the GT-based LCV. Our findings demonstrate that the collision probability of the SDP-based LCV has high safety with no collisions, which is significantly lower than the collision rate of the GT-based LCV of around 6.6%. In the harsh conditions, the GT-based LCV experiences a notable increase in collision rate, approximately 15.3%, while the SDP-based LCV demonstrates a minimal increase of only about 0.9%. These results highlight the robust performance for safety of the proposed decision-making even under harsh conditions. Moreover, the longitudinal jerk of the SDP-based LCV exhibited lower maximum, minimum, and variance values when compared to the GT-based LCV. Consequently, the proposed algorithm for LCV is capable of making decisions that are both safer and more comfortable.

To thoroughly evaluate effectiveness of the proposed algorithm, two distinct tests Test 1 and Test 4 are compared. The results of both cases are presented in Table III, which includes the average velocity, minimum and maximum longitudinal jerk, collision rate, and average distance between LKV and LCV. Our proposed algorithm exhibits good performance in terms of safe and comfortable decision-making, as evidenced by lower minimum and maximum longitudinal jerk values, and a reduced collision rate when compared to the IDM-LKV and GT-LCV scenario, which has a high collision rate of approximately 61%. Furthermore, the collision rate and minimum and maximum of longitudinal jerk increased compared to Tests 2 and 3. This observation indicates that both IDM and GT exhibit safer and more comfortable decision-making while driving with the proposed algorithm based on SDP. Consequently, the proposed method has excellent performance in terms of safety and comfort. In addition, the proposed algorithm shows a higher average distance and velocity, which indicates that our decision-making approach can effectively control vehicle with high velocity while maintaining a safe distance between them.

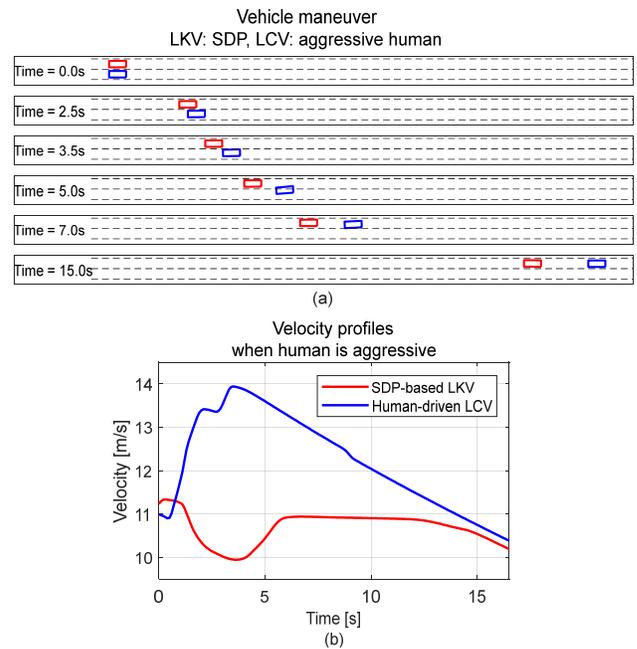

**Fig. 15.** Results of decision-making when the LKV drives based on the proposed method with aggressive human-driven LCV (a) maneuver of two vehicles and (b) velocity profiles.

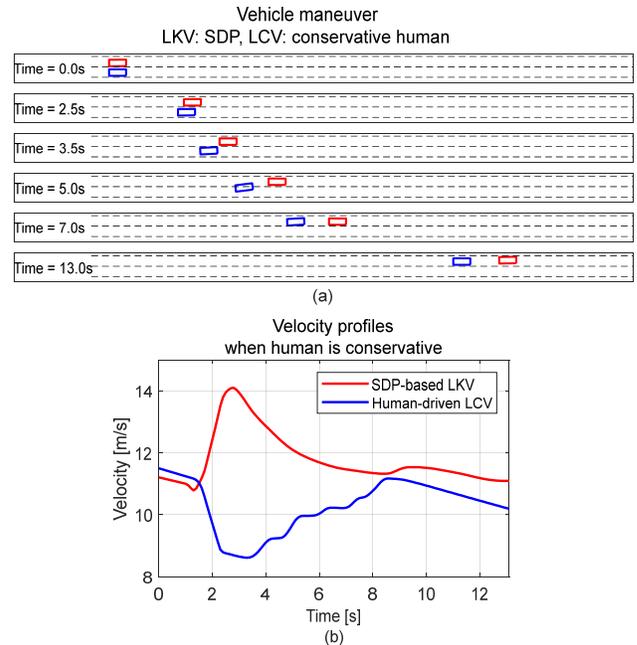

**Fig. 16.** Results of decision-making when the LKV drives based on the proposed method with conservative human-driven LCV (a) maneuver of two vehicles and (b) velocity profiles.

## V. EXPERIMENTAL VALIDATION

As autonomous vehicles become commercialized and share the roads with human-driven vehicles, it is crucial for them to navigate safely and efficiently through interactions with a variety of human drivers, including aggressive and conservative ones. In Section IV, we present a validation of our proposed decision-making algorithm for autonomous vehicles driving among other autonomous vehicles. In this section, we evaluate the performance of our cooperative and interaction-aware decision-making approach when an autonomous vehicle using our proposed method shares the road with a human-driven vehicle.

### A. Experimental Setup

Evaluating the performance of decision-making algorithm for autonomous vehicles where the interaction between vehicles is occurred needs a wide area of roads with well controlled environment, which requires high cost and large engineering effort. In addition, repeated lane change scenarios have high risk of accidents. Therefore, evaluating the performance of the autonomous vehicles in actual road driving conditions is almost impossible.

For human-driven and autonomous vehicles to coexist on the same road, a driving platform that accommodates both vehicles is necessary. In this study, such a driving platform is implemented in a networked traffic simulator, as shown in Fig. 14. Each vehicle is emulated in an independent driving simulator and can be controlled by either a human driver or an autonomous vehicle's decision-making algorithm. The driving simulators are connected through Ethernet communication, and the vehicle physics model for each vehicle is implemented using the widely utilized Carsim software for vehicle dynamics studies. Both vehicles can be simultaneously driven on the same road, with all motions synchronized across the two simulators.

### B. Results

To evaluate the proposed decision-making for the LKV, a human drives the LCV, and experiments are conducted in two cases: the human is aggressive and conservative. In both cases, the same policy of LKV is used, and the results are shown in the Fig. 15 and 16.

The presented results illustrate the maneuver and the velocity profiles of two vehicles. Specifically, Fig. 15 shows the results of driving with an aggressive human-driven LCV and Fig. 16 demonstrates the results of driving with a conservative human-driven LCV. When the aggressive human changes the lane through acceleration, the proposed method-based LKV cooperatively and interactively drives through deceleration. On the other hand, when the conservative human decelerates and changes the lane, the proposed method-based LKV effectively drives through acceleration.

To evaluate the proposed decision-making for the LCV, a human is driver of LKV, and experiments are conducted in two cases: the human is aggressive and conservative. In both cases, the same policy of LCV is used, and the results are shown in the Fig. 17 and 18.

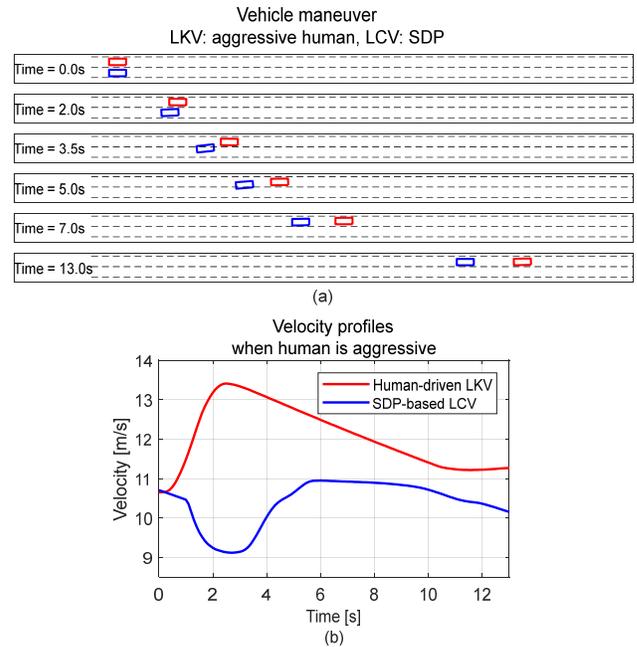

**Fig. 17.** Results of decision-making when the LCV drives based on the proposed method with aggressive human-driven LKV (a) maneuver of two vehicles and (b) velocity profiles.

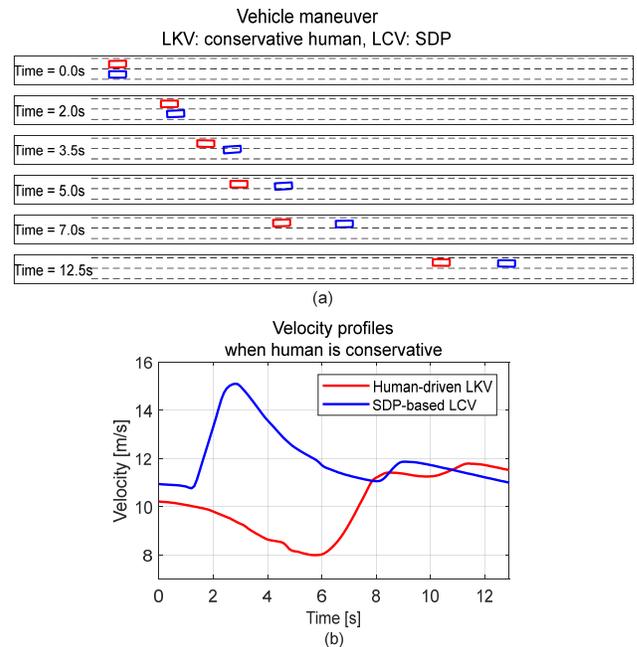

**Fig. 18.** Results of decision-making when the LCV drives based on the proposed method with conservative human-driven LKV (a) maneuver of two vehicles and (b) velocity profiles.

The presented results illustrate the maneuver and the velocity profiles of two vehicles. Specifically, Fig. 17 shows results of driving with an aggressive human and Fig. 18 demonstrates the results of driving with a conservative human. When an aggressive human driver exhibits an intention not to yield through acceleration, our proposed algorithm-based LCV method cooperatively and interactively changes lanes through deceleration. Conversely, in cases where the conservative human driver displays an intention to yield through deceleration, the proposed method-based LCV effectively changes lanes through acceleration. Therefore, these findings confirm that the proposed decision-making algorithm for LKV and LCV can cooperatively and effectively drive with human-driven vehicles with various characteristics while reflecting implicit interactions.

## VI. CONCLUSION

This study proposed a cooperative and interaction-aware decision-making algorithm for autonomous vehicles, specifically in lane change scenarios. The algorithm stochastically considers the future behavior of other vehicles based on actual driving data and defines an interaction model based on relative information between vehicles with fewer states. The proposed algorithm considers safety, comfort, intention, and character of the ego vehicle, and is developed for both LKV and LCV.

To validate the cooperative and interactive driving, simulation results with vehicles of various driving styles, such as aggressive and conservative, demonstrated that the proposed algorithm enables interactive and cooperative driving. Moreover, the proposed algorithm enables safe and comfortable driving with lower collision and longitudinal jerk compared to the IDM and GT-based methods commonly used in studies. Furthermore, the proposed algorithm demonstrates the ability to drive cooperatively and interactively with drivers of diverse characteristics.

Given the potential scenario of commercialized autonomous vehicles, we have verified the proposed method for driving with human-driven vehicles. Our proposed method enables cooperative driving with human-driven LKV and LCV of varying characteristics, while effectively reflecting implicit interactions.

In summary, the proposed cooperative and interaction-aware decision-making algorithm presents a promising approach for safe and efficient driving of autonomous vehicles. This study's results demonstrate the algorithm's effectiveness in accommodating various driving styles and enabling comfortable, safe, and interactive driving. The proposed method can be easily implemented in real-time and extended beyond lane change to other driving scenarios. In addition, the proposed method generates a variety of decision-making models and can be operated with human drivers. Thus, it has the potential to be extended and applied to the development of human-like autonomous vehicles.

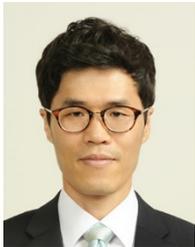

**Changsun Ahn** received the B.S. and M.S. degrees from Seoul National University, Seoul, South Korea, in 1999 and 2005, respectively, and the Ph.D. degree from the University of Michigan, Ann Arbor, in 2011, all in mechanical engineering. He is currently a Professor at Pusan National University, South Korea. His research interests include the fields of automotive control/estimation. Recently, he is focused on autonomous vehicle control and hybrid vehicle control.

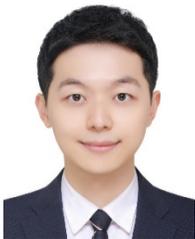

**Jemin Woo** received the B.S. degree from Pusan National University, Busan, South Korea, in 2018, in mechanical engineering. He is currently pursuing the Ph.D. degree in mechanical engineering from Pusan National University, Busan, South Korea. His research interests include automotive system modeling and autonomous vehicle control. Recently, he is focused on driver model design for autonomous vehicles.